\documentclass[5p,number,sort&compress]{elsarticle}

\usepackage{amsmath,amssymb,pstricks,dcolumn,booktabs}
\usepackage{hyperref}
\usepackage[load-configurations = version-1]{siunitx}
\usepackage{lineno}

\newcommand{\tamutrap}{\texttt{T\small AMUTRAP}}

\makeatletter
\def\ps@pprintTitle{%
  \let\@oddhead\@empty
  \let\@evenhead\@empty
  \def\@oddfoot{\reset@font\hfil\thepage\hfil}
  \let\@evenfoot\@oddfoot
}
\makeatother

\begin{document}

\begin{frontmatter}

\title{Design of a Unique Open-Geometry Cylindrical Penning Trap}
\author[CI,PHYS]{M. Mehlman\corref{cor1}}
\cortext[cor1]{Corresponding author, email: mehlmanmichael@tamu.edu, phone: (979) 845 1411, fax: (979) 845 1899}
\author[CI]{P.D. Shidling\corref{}}
\author[CI,CHEM]{R.S. Behling\corref{}}
\author[CI,PHYS]{B. Fenker\corref{}}
\author[CI,PHYS]{D. Melconian\corref{}}
\author[ENG]{L. Clark\corref{}}
\address[CI]{Cyclotron Institute, Texas A\&M University, 3366 TAMU, 
  College Station, TX, 77843-3366}
\address[PHYS]{Department of Physics and Astronomy, Texas A\&M University, 
  4242 TAMU, College Station, TX, 77843-4242}
\address[CHEM]{Department of Chemistry, Texas A\&M University, 
  3012 TAMU, College Station, TX, 77842-3012}
\address[ENG]{College of Engineering, Texas A\&M University, 
  3126 TAMU, College Station, TX, 77843-3126}

\date{today}

\begin{abstract}
  The Texas A\&M University Penning Trap facility is an upcoming 
  ion trap that will be used to search for possible scalar currents in 
  $T=2$ superallowed $\beta$-delayed proton decays, which, if found, would be 
  an indication of physics beyond the standard model.  In addition, \tamutrap{} 
  will provide a low-energy, point-like source of ions for various other 
  applications at the Cyclotron Institute. The experiment is centered around 
  a unique, compensated cylindrical Penning trap that employs a specially 
  optimized $\mathrm{length}/\mathrm{radius}$ ratio in the electrode structure 
  that is not used by any other facility.  This allows the geometry to exhibit 
  an unprecedented 90~mm free radius, which is larger than in any existing 
  trap, while at the same time remaining a tractable overall length. The trap 
  geometry was designed from first principles to be suitable for a wide range 
  of nuclear physics experiments.  In particular, the electrode structure is 
  both ``tunable'' and ``orthogonalized'', which allows for a near quadrupole 
  electric field at the trap center, a feature necessary for performing 
  precision mass measurements.
\end{abstract}
\begin{keyword}
  Penning \sep cylindrical \sep precision \sep analytic solution
  \PACS 23.40.bw \sep 24.80.+y \sep 37.10.Ty \sep 07.75.+h
\end{keyword}

\end{frontmatter}


\section{Introduction}
Low energy precision $\beta$-decay experiments have proven to be an 
excellent compliment to high energy physics for placing new constraints 
on physics beyond the standard model (SM)~\cite{Severijns2011, Severijns2006, 
  Behr2009}.  Up to this point, it has been possible to explain the results 
from such experiments by a time reversal invariant $V-A$ interaction 
displaying a maximal violation of parity; however, more precise measurements 
of $ft$ values~\cite{Hardy2010,Severijns2002} and correlation 
parameters~\cite{Adelberger1999} in particular $\beta$-decays can serve to 
test for the presence and properties of any non-SM processes that may occurr 
in such interactions.

\subsection{Motivation}
\begin{figure}
  \centering\includegraphics[width=\linewidth]{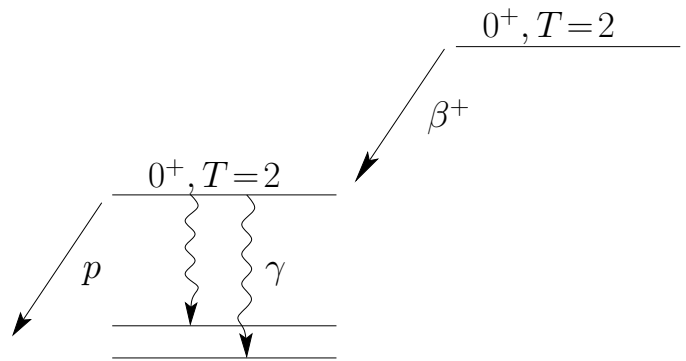}
  \caption{Decay scheme for a generic $T=2$, $0^{+}\rightarrow0^{+}$ 
    superallowed $\beta$-delayed proton decay.\label{decayscheme}}
\end{figure}

The initial experimental program at the upcoming Texas A\&M University 
Penning Trap (\tamutrap) facility will seek to improve the limits 
on non-SM processes in the weak interaction, in particular scalar currents, 
by measuring the $\beta-\nu$ correlation parameter, $a_{\beta\nu}$, for $T=2$, 
$0^{+}\rightarrow0^{+}$ \mbox{superallowed} $\beta$-delayed proton emitters 
(the generic decay scheme is shown in Fig.~\ref{decayscheme}, and the 
preliminary list of nuclei to be studied is outlined in Table~\ref{t2nuclei}).

The general $\beta$-decay rate with no net polarization or alignment is given 
by \cite{Jackson1957}:
\begin{equation}\label{tdecaycrosssection}
  \frac{d^{5}\Gamma}{dE_ed\Omega_ed\Omega_\nu}\propto 1 + 
  \frac{p_e}{E_e}a_{\beta\nu}\cos{\theta_{\beta\nu}} + b\frac{m_e}{E_e},
\end{equation}
where, $E_e$, $p_e$, and $m_e$ are the energy, momentum, and mass of the 
$\beta$, $\theta_{\beta\nu}$ is the angle between the $\beta$ and $\nu$, and 
$b$ is the Fierz interference coefficient.  Thus, it is possible to determine 
the $\beta-\nu$ correlation parameter by means of an experimental measurement 
of the angular distribution between the $\beta$ and $\nu$.  For the strict 
$V-A$ interaction currently predicted by the SM (in pure Fermi decays) the 
$\beta-\nu$ angular distribution should yield a value for $a_{\beta\nu}$ of 
exactly 1.  Any admixture of a scalar current to the predicted interaction, 
a result of particles other than the expected $W^\pm$ being exchanged during 
the decay, would result in a measured value of $a_{\beta\nu}<1$.

\begin{table}
  \centering
  \sisetup{tabformat=3.2,tabnumalign=center,tabunitalign=center,tabtextalign=left}
  \begin{tabular}{SSSS}
    \toprule
    {Nuclide} & {Lifetime (ms)} & 
    {$E_p$ (MeV)} & {$R_L$ (mm)}\\
    \midrule
    {$^{20}$Mg} & 137.05 & 4.28 & 42.7\\
    {$^{24}$Si} & 147.15 & 3.91 & 40.8\\
    {$^{28}$S}  & 180.33 & 3.70 & 39.7\\
    {$^{32}$Ar} & 141.38 & 3.36 & 37.8\\
    {$^{36}$Ca} & 141.15 & 2.55 & 33.0\\
    {$^{40}$Ti} & 72.13 & 3.73 & 39.9\\
    {$^{48}$Fe} & 63.48 & 1.23 & 22.9\\
    \bottomrule
  \end{tabular}
  \caption{The $T=2$ nuclei that will compose the initial experimental program 
    measuring $a_{\beta\nu}$.  The Larmour radii, $R_L$, for the ejected 
    protons of interest (having energy $E_p$) shown are calculated for the 
    7~T magnetic field of \tamutrap. \label{t2nuclei}}
\end{table}

\tamutrap{} will observe this angular distribution between $\beta$ and $\nu$ for 
$\beta$-delayed proton emitters in order to take advantage of the benefit 
these particular decays have on the experimental procedure.  In such a case, 
the $\beta$-decay yields a daughter nucleus that is unstable, and can result 
in the subsequent emission of a proton with significant probability.  As 
discussed in~\cite{Adelberger1999}, the great advantage to utilizing 
$\beta$-delayed proton emitters for such a study is that this proton 
energy distribution contains information about $\theta_{\beta\nu}$.  If 
the $\beta$ and $\nu$ are ejected from the parent nucleus in the same 
direction, they will impart a larger momentum kick to the daughter nucleus, 
which will be inherited by the proton.  Conversely, if the $\beta$ and $\nu$ 
are emitted in opposite directions, this momentum kick is reduced.  By 
measuring the proton energy distribution at \tamutrap{} the value of 
$a_{\beta\nu}$ will be deduced, which can then indicate the existence of 
scalar currents in these decays~\cite{Jackson1957}.

\subsection{Cylindrical Traps\label{cylindricalsection}}
A cylindrical Penning trap geometry allows for efficient access to the 
trapped ions, a large trapping volume compared to a hyperboloid trap 
geometry, and an electric field that can be described analytically, which 
is of particular importance during the design process.  Additionally, 
cylindrical electrodes are more easily manufactured with higher precision.  
For these reasons, Penning traps with a cylindrical geometry have been widely 
employed in nuclear physics research experiments ranging from precision mass 
measurements~\cite{Blaum2006} to the production of 
anti-hydrogen~\cite{Amoretti2002}.

Precision $\beta$-decay experiments are well served by a cylindrical geometry 
due to the fact that the magnetic field employed to trap the ions radially 
may simultaneously be used to contain charged decay products~\cite{Kozlov2006} 
such as $\beta$'s and protons with up to $4\pi$ acceptance in an appropriately 
designed trap.  The strong magnetic radial confinement in combination with the 
weak electrostatic axial confinement direct the decay products of interest to 
either end of the trap for detection with negligible affect on the energy of 
the particles.  At the same time, features of a cylindrical trap geometry 
can be useful for other nuclear physics experiments, such as maintaining a 
line of sight to the trap center for spectroscopy, an easily ``tunable'' and 
``orthogonalized'' electric field for experiments requiring a harmonic
 potential (such as mass measurements), and unrivaled access to the trapped 
ions.

\section{Design of \tamutrap}
For the reasons mentioned in~\S\ref{cylindricalsection}, a cylindrical 
geometry has been chosen for the \tamutrap{} measurement Penning 
trap~\cite{Mehlman2012}.  This particular geometry has been optimized to 
create a design that is suitable both for the precision $\beta$-decay 
experiments of interest, as well as a wide range of nuclear physics 
experiments as discussed.  Specifically, the trap must display a large-bore 
for containment of decay products, it should allow for the placement of 
biased detectors at both ends for observation of these products, and it 
should exhibit a ``tunable'' and ``orthogonalized'' geometry in order to 
achieve a harmonic electric field.

\subsection{Large-bore}
For $\beta-\nu$ correlation measurements, the trap must have a free diameter 
large enough to contain the decay products of interest within the electrodes 
via the Lorentz force imposed by the trapping magnetic field.  The initial 
program of measuring $a_{\beta\nu}$ will investigate the $T=2$ nuclei shown 
in Table~\ref{t2nuclei}, by observing the proton energy distribution.  To 
fully contain protons of interest with full $4\pi$ acceptance the trap radius 
is set to twice the Larmour radius of the most energetic expected proton 
within the 7~T magnetic field provided by the Agilent 7T 210 ASR 
magnet~\cite{Agilent}.  The trap radius was chosen to be 90 mm, which will 
fully contain protons of up to 4.75 MeV.  This radius will be the largest of 
any existing Penning trap and will easily contain the protons of interest, as 
well as the less magnetically rigid $\beta$'s, for the initial $a_{\beta\nu}$ 
studies.

\subsection{Endcaps}
The other primary requirement for performing the mentioned correlation 
measurements is that the design must ultimately accommodate position sensitive 
strip detectors at either end of the trap.  The charged decay products 
($\beta$, $p$, daughter) exhibit Larmour precession contained completely 
within the bore of the Penning trap until they are detected by such a 
detector.  These detectors have been simulated separately as disk-shaped 
``endcap electrodes.''  Such an approximation satisfies the need to bias 
the detectors at some arbitrary potential not necessarily equal to that of 
the hollow cylindrical end electrodes.  The impact of the endcap electrodes 
on the solution to the complete electric field is discussed in detail 
in~\S\ref{tunablesection}.

\subsection{Tunable\label{tunablesection}}
A priority for the \tamutrap{} facility is to attain a very good quadrupole field 
at the center of the measurement Penning trap.  To achieve this, the trap 
design must be ``tunable'', that is, it must make use of compensation 
electrodes that serve to adjust the field shape.  Other cylindrical 
traps~\cite{Roux2011,Beck1997} already employ tunable geometries; however, 
these configurations were not suitable for \tamutrap{} due to the large-bore 
requirement: enlarging such geometries (which does not inherently affect the 
field shape if all features are scaled appropriately), results in a trap too 
long to fit within the available 7~T magnet.  In addition, the analytic 
solutions of the electric field used to design these existing 
traps~\cite{Gabrielse1989} directly employ long-endcap approximations, which 
makes a new design utilizing the existing field calculations not applicable 
to the short-endcap / large-bore requirement necessitated by the envisioned 
\tamutrap{} experimental program.

\begin{figure}
  \centering\includegraphics[width=\linewidth]{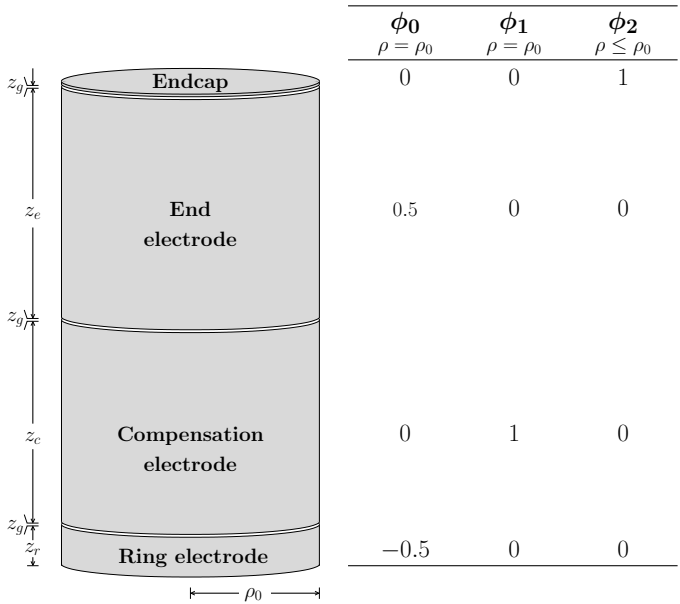}
  \caption{The trap geometry and boundary conditions used in the analytic 
    solution.\label{boundaryconditions}}
\end{figure}

For these reasons, a new analytic solution for a short-endcap, tunable Penning 
trap was derived from first principles, in part following the discussion 
in Ref.~\cite{Gabrielse1989}.  The contribution to the potential due to the 
various electrodes (Fig.~\ref{boundaryconditions}) can be found by noting that 
any potential may be expanded in terms of the Legendre polynomials, $P_k$, 
the potential depth, radius from the trap axis, and characteristic trap 
distance~\cite{Gabrielse1989} , $V_0$, $r$, and $d$, respectively, and the 
expansion coefficients, $C_k$ (here $k$ is even due to symmetry across the 
trapping plane):
\begin{align}
  V&=\frac{1}{2}V_0\sum_{\substack{k=0\\k\ \mathrm{even}}}^\infty C_k
  \left(\frac{r}{d}\right)^kP_k(\cos{\theta}),\label{tlegendre}
  \intertext{with}
  d&=\sqrt{\tfrac{1}{2}\left[(z_{r}+z_{g}+z_{c}+z_{g})^2+
      \tfrac{1}{2}\rho_0^2\right]}.\label{chartrapd}
\end{align}
By superposition, the potential at the trap center may also be written as a 
sum of the potentials of each of the contributing electrodes:
\begin{equation}\label{tsuperposition}
  V=V_0\phi_0+V_1\phi_1+V_2\phi_2.
\end{equation}
Here, $\phi_0$ is due to the boundary conditions generated by the ring and 
end electrodes (the primary potential well), and $\phi_1$ and $\phi_2$ 
are from the compensation and endcap (detector) electrodes respectively 
(Fig.~\ref{boundaryconditions}), while the $V_i$'s are the potentials at 
which these electrodes are held.  Each of the $\phi$'s can in turn be expanded 
in Legendre polynomials and substituted back into Eq.~\ref{tsuperposition}, 
which yields the simple result:
\begin{equation}\label{tsimpleresult}
  C_k=D_k+E_k\frac{V_1}{V_0}+F_k\frac{V_2}{V_0},
\end{equation}
where $D_k$, $E_k$, and $F_k$ are the individual expansion coefficients 
due to the ring and end electrodes, compensation electrodes, and endcap 
(detector) electrodes respectively, and must be solved for individually.  
$D_k$ and $E_k$, which are due to hollow cylinder shaped electrodes, can 
be found by expanding $\phi_0$ and $\phi_1$ in Bessel functions, 
$J_\alpha$.  After applying the periodic boundary condition in $z$, 
$\phi(z)=\phi(-z)$, it can be found that 
\begin{equation}\label{tbesselfunction}
  \phi_i=\sum_{n=0}^\infty A_nJ_0(\imath k_n\rho)
  \cos{(k_nz)},
\end{equation}
where $A_n$ is an additional expansion coefficient.  Here $k_n$ is due to 
the periodic boundary condition, and is given by
\begin{equation}\label{tkn}
  k_n=\frac{(n+\frac{1}{2})\pi}{z_\mathrm{tot}},
\end{equation}
where $z_\mathrm{tot}=z_{r}+z_{c}+z_{e}+3z_{g}$.  Setting 
Eq.~\ref{tbesselfunction} equal to the expansion in Legendre polynomials 
(Eq.~\ref{tlegendre}) allows one to solve for the expansion coefficients of 
the Legendre polynomials by equating the two along $\hat{z}$.  This results 
in the following solutions:
\begin{align}
  D_k & =\sum_{n=0}^\infty\frac{2A_n^Dd^kk_n^k(-1)^{k/2}}{k!}
  \nonumber\\[-0.65em]
  \\[-0.65em]
  E_k & =\sum_{n=0}^\infty\frac{2A_n^Ed^kk_n^k(-1)^{k/2}}{k!}.\nonumber
\end{align}
The coefficients $A_n^i$ are subsequently determined by applying the 
appropriate boundary conditions (Fig.~\ref{boundaryconditions}) along with 
the orthogonality of cosine, yielding
\begin{align}
  A_n^D & =\frac{(-1)^n - \sin[k_n(z_r+z_g+z_c+z_g)] - \sin(k_nz_r)}
  {k_nz_\mathrm{tot}J_0(\imath k_n\rho_0)}\nonumber\\[-0.5em]
  \label{tanc4}\\[-0.5em]
  A_n^E & =\frac{2\big(\sin[k_n(z_r+z_g+z_c)] - \sin[k_n(z_r+z_g)]\big)}
  {k_nz_\mathrm{tot}J_0(\imath k_n\rho_0)}.\nonumber
\end{align}
The contribution to the potential from the endcap electrodes must be handled 
differently.  Here $\phi_2$ is defined at $z_\mathrm{tot}$ for any radius 
less than $\rho_0$.  After simplifying due to the cylindrical symmetry of 
the system, $\phi_2$ can be written:
\begin{equation}\label{tphi23}
  \phi_2=\sum_{n=0}^{\infty}J_{m}(k_{0n}\rho)e^{k_{0n}z}B_n,
\end{equation}
where, $k_{mn}$ is related to $x_{mn}$, the $n$\textsuperscript{th} zero of 
the $m$\textsuperscript{th} Bessel function as in
\begin{equation}\label{tphi22}
  k_{mn}=\frac{x_{mn}}{\rho_0}.
\end{equation}
$B_n$ can now be determined through the application of appropriate boundary 
conditions (see Fig.~\ref{boundaryconditions}) and the Bessel function 
orthogonality relation, giving
\begin{equation}\label{tphi26}
  B_n = \frac{2e^{-k_{0n}z_\mathrm{tot}}}{x_{0n}J_{1}(x_{0n})}.
\end{equation}
Taking into account both endcaps (at $\pm z$) yields the complete formulation 
of the potential due to the endcaps at any $z$:
\begin{multline}\label{tphi27}
  \phi_2 = \frac{V_2}{2}\left[ 
    \sum_{n=1}^{\infty}J_0\left(\frac{x_{0n}}{\rho_0}
      \rho\right)e^{z\frac{x_{0n}}{\rho_0}}
    \frac{2e^{-k_{0n}z_\mathrm{tot}}}{x_{0n}J_{1}(x_{0n})}\right. \\
  + \left.\vphantom{\sum_{n=1}^{\infty}J_0}
    \sum_{n=1}^{\infty}J_0\left(\frac{x_{0n}}{\rho_0}
      \rho\right)e^{-z\frac{x_{0n}}{\rho_0}}
    \frac{2e^{-k_{0n}z_\mathrm{tot}}}{x_{0n}J_{1}(x_{0n})}
  \right].
\end{multline}
This result for the potential was subsequently Taylor expanded using 
\texttt{Mathematica}~\cite{Mathematica}, yielding coefficients which may be 
referred to here by $T_k$.  Setting the result from the expansion equal to 
the potential expanded in Legendre polynomials along the $z$-axis and equating 
terms yields the final result for the coefficients $F_k$:
\begin{equation}\label{tphi211}
  F_k=V_0T_kd^k.
\end{equation}
All expansion coefficients from Eq.~\ref{tsimpleresult} have now been 
completely defined, and the electric field at the trap center can be 
specified to arbitrary precision.  By construction, it is easy to 
characterize the components of this field, which allows for a straightforward 
optimization.

\subsection{Orthogonal}
For certain experiments (such as precision mass measurements), it is crucial 
to be able to tune out the anharmonic terms ($C_{\ge4}\rightarrow0$) of the 
electric field during the course of a measurement without affecting the 
harmonic ($C_2$) component of the field.  In order to achieve this for 
\tamutrap, the procedure discussed in Ref.~\cite{Gabrielse1989} was followed.

The strength of the dominant anharmonic term for the superposition of the 
electric fields is given by the coefficient $C_4$, and the second most 
dominant contribution comes from $C_6$, which has an affect on the shape 
of the electric field smaller than $C_4$ by a magnitude of ${(r/d)}^{2}$ 
(where $r$ is that radius of ion motion and $d$ is the characteristic trap 
dimension)~\cite{Gabrielse1989}.  These coefficients are determined, in turn, 
by the anharmonic contributions from the constituent electrodes, that is, 
$D_4$, $D_6$, $E_4$, $E_6$, $F_4$, and $F_6$.  $C_4$ may always 
be tuned out by adjusting the potential on the compensation electrodes until 
the field is essentially harmonic.  For a general geometry, however, this 
procedure will affect the value of $C_2$.  Since only the potentials of 
the compensation electrodes are adjusted in this process, it is possible to 
eliminate this affect by requiring that $E_2=0$, that is that the 
compensation electrodes have no influence on the harmonic term of the 
superposition, $C_2$.

The expansion coefficient $E_2$, which is a function of the entire geometry, 
was minimized using the analytic solution derived above.  To do this, the 
constraints imposed by other considerations (trap bore, assembly 
considerations, etc.) were first imposed.  This left three free parameters: 
ring electrode length, compensation electrode length, and end electrode 
length.  Ring electrode length and end electrode length were chosen in order 
to both minimize $C_6$ and achieve a large tunability with respect to 
$C_4$, where tunability is described in Ref.~\cite{Gabrielse1983}:
\begin{equation}\label{tunability}
  \text{Tunability}=V_0\frac{dC_4}{dV_1}.
\end{equation}
After determining the ring and end electrode lengths in this way, $E_2$ was 
minimized with respect to the remaining parameter, the compensation electrode 
length, thereby ``orthogonalizing'' the geometry.  Since $E_2$ changes sign 
when scanning over electrode length, it was possible to choose a geometry 
that resulted in an arbitrarily small value for $E_2$.
\section{\tamutrap\label{TAMUTRAP}}
\begin{figure}
  \centering\includegraphics[width=\linewidth]{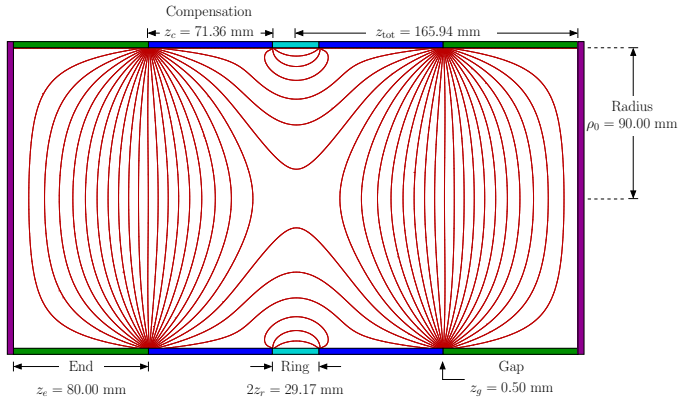}
  \caption{The optimized trap geometry with \texttt{SIMION} generated electric 
    fields lines overlayed.\label{finalgeometry}}
\end{figure}

Performing the optimization as discussed above resulted in the geometry shown 
in Fig.~\ref{finalgeometry}.  The trap radius is 90~mm, which is larger than 
in any existing Penning trap.  The ring electrode length is 29.17~mm, the 
compensation electrode length is 71.36~mm, the end electrode length is 
80.00~mm, and 0.50~mm gaps have been accounted for.  This geometry results in a 
$\mathrm{length}/\mathrm{radius}$ ratio of $l/r_0=3.72$.  A good quadrupole 
field ($C_4=-6.8\times10^{-6}$, $C_6=6.2\times10^{-6}$) has been 
calculated to be achievable with compensation electrodes set to 
$V_1=-0.373V_0$ (where $V_0$ is the primary trap depth).  The analytic 
expansion of the electric field around the trap center up to $C_8$ is 
shown in Table~\ref{expansioncoefficients}.  Changing the voltage on the 
endcap electrodes (detectors) will adjust the predicted tuning (compensation) 
voltage; however, the trap will always remain ``tunable'' and 
``orthogonalized'' since the contributions to the potential for each 
electrode are independent by superposition.

\section{Simulation}
The analytic solution for the proposed geometry was verified using 
\texttt{SIMION}~\cite{Simion}, an electric field and ion trajectory simulation 
program, in order to confirm the validity of the calculations.  The results 
from \texttt{SIMION} are listed in Table~\ref{expansioncoefficients} (along with the 
analytic solutions from~\S\ref{TAMUTRAP}), and the resulting equipotential 
lines have been overlayed on the geometry cross section presented in 
Fig.~\ref{finalgeometry}.  The values output by \texttt{SIMION} agree with the analytic 
solutions for each $C_k$ to a few parts in $10^{-3}$.  The discrepancies 
between the analytic and simulated values can be accounted for by the 
inherent pixelation of the geometry as represented in \texttt{SIMION} and processing 
constraints, which have both been minimized as far as allowed by RAM and 
available computing time.  Specifically, the optimized geometry has been 
represented in the simulation to the nearest $0.01$~mm, and the voltages 
have been defined to $1\times10^{-5}$~V.

\begin{table}
  \sisetup{tabformat=2e3,tabnumalign=center,expproduct=tighttimes}
  \centering\begin{tabular}{cSSSS}
    \toprule
    {$C_i$} & {TAMU} & {TAMU} & {TITAN}& {LEBIT} \\
    {} & {Analytic} & {Simulated} & {Analytic} & {Simulated}\\
    \midrule
    {$C_0$} & -5e-1 & -5e-1 & {{-}} & 8e-1\\
    {$C_2$} & +5e-1 & +6e-1 & {{-}} & 1e0\\
    {$C_4$} & -7e-6 & +9e-4 & -7e-6 & 2e-3\\
    {$C_6$} & +6e-6 & -3e-3 & +5e-5 & -4e-3\\
    {$C_8$} & -4e-2 & -4e-2 & {{-}} &  3e-3\\
    \bottomrule
  \end{tabular}
  \caption{Expansion coefficients are compared for the optimized \tamutrap{} 
    measurement trap when tuned (analytic and simulated) and two other 
    existing Penning traps: TITAN (calculated analytically as in 
    Ref.~\cite{Brodeur2010}), and LEBIT (simulated using \texttt{SIMION} as in 
    Ref.~\cite{Ringle2009}).\label{expansioncoefficients}}
\end{table}

\section{Comparison to Existing Traps}
The defining features of the \tamutrap{} geometry described here are the unique 
inner radius (90~mm), the small $\mathrm{length}/\mathrm{radius}$ ratio 
($l/r_0=3.72$), and the consideration made for detectors (short endcap 
electrodes).  This new $l/r_0$ ratio is what allows \tamutrap{} to exhibit 
such an unprecedented radial size when compared to other cylindrical Penning 
traps, while still measuring only 335~mm in overall length.  The optimized 
ISOLTRAP geometry, for example, has an $l/r_0=
11.75$~\cite{Raimbault-Hartmann1997}, and would therefore require over 1~m in 
length in order to maintain geometrical proportions (and, therefore, electric 
field shape) with a 90~mm inner radius.  The compact geometry employed by 
\tamutrap{} allows for a structure with a very large radius to easily fit 
within the 1-m long bore of the 7T 210 ASR magnet.  At the same time, the new 
analytic solution described in~\S\ref{tunablesection} retains the good 
quadrupole nature of the electric field displayed by other prominent Penning 
traps, which is required for other experiments such as precision mass 
measurements. 

Table~\ref{expansioncoefficients} compares the analytic and simulated electric 
field expansion coefficients of \tamutrap{} to an analytic solution reported by 
LEBIT~\cite{Ringle2009}, and a simulated solution reported by 
TITAN~\cite{Brodeur2010}.  The suppression of the anharmonic terms in the 
electric field generated by the geometry for the \tamutrap{} measurement 
Penning trap is comparable to that presented by these two prominent 
mass-measurement facilities, for which a very well-tuned harmonic electric 
field is critical~\cite{Gabrielse2009}.  With respect to the inherent field 
shape, the \tamutrap{} geometry should therefore be suitable for such 
precision mass measurements; however, it remains to be seen what affects the 
unprecedented electrode size and trapping volume necessitated by the primary 
program of performing $\beta-\nu$ correlation measurements will have on the 
specific procedures required in these studies.  In particular, due to the 
enlarged geometry, the uniformity of the applied potentials across the 
electrodes is likely to be worse at \tamutrap{} than at existing facilities 
that employ smaller electrodes.  It remains to be seen, however, whether or 
not these effects will ultimately limit mass resolution.  In either case, 
they will not have a significant effect on the main program where the harmonic 
potential is not as stringent.

\section{Conclusion}
A novel, large-bore cylindrical Penning trap with a new 
$\mathrm{length}/\mathrm{radius}$ ratio in the electrode structure has been 
described.  The design work has been performed with the initial research 
program of measuring correlation parameters for $T=2$ superallowed 
$\beta$-delayed proton emitters in mind; however, careful attention has also 
been paid to creating a facility with maximum suitability for a wide range of 
possible future nuclear physics experiments.  In particular, a line of sight 
to the trap center, an open geometry, and a well characterized quadrupole 
electric field are predicted with the proposed design.  The analytic solutions 
to the electric field were checked against simulation, and match to high 
precision.  These values, which are comparable to those presented by existing 
mass measurement facilities, suggest that \tamutrap{} should be capable of 
performing mass measurements in addition to its primary program.  The 
aforementioned features, in conjunction with the unprecedented 90~mm free 
trap radius, will make the \tamutrap{} facility a unique tool for future 
nuclear physics research.

\section{Acknowledgments}
The authors would like to thank Guy Savard, Jason Clark, and Jens Dilling for 
fruitful discussions and advice.  This work was supported by the U.S. 
Department of Energy under Grant Numbers ER41747 and ER40773.

\bibliographystyle{elsarticle-num}
\bibliography{references}

\end{document}